 \documentclass[preprints,article,accept,moreauthors]{Definitions/mdpi}
\firstpage{1} 
\pubvolume{xx}
\issuenum{1}
\articlenumber{5}
\pubyear{2019}
\copyrightyear{2019}
\history{Received: date; Accepted: date; Published: date}

\usepackage{amsmath}
\usepackage{amsthm}
\usepackage{tikzfig}
\usepackage{amsfonts}

\tikzstyle{env}=[copoint,regular polygon rotate=0,minimum width=0.2cm, fill=black]

\tikzstyle{every picture}=[baseline=-0.25em]
\tikzstyle{dotpic}=[scale=0.5]
\tikzstyle{diredges}=[every to/.style={diredge}]
\tikzstyle{dot graph}=[shorten <=-0.1mm,shorten >=-0.1mm,scale=0.6]
\tikzstyle{plot point}=[circle,fill=black,minimum width=2mm,inner sep=0]

\tikzstyle{braceedge}=[decorate,decoration={brace,amplitude=2mm,raise=-1mm}]
\tikzstyle{small braceedge}=[decorate,decoration={brace,amplitude=1mm,raise=-1mm}]
\tikzstyle{left hook arrow}=[left hook-latex]
\tikzstyle{right hook arrow}=[right hook-latex]

\tikzstyle{dtriangle}=[fill=yellow,draw=black,shape=isosceles triangle,shape border rotate=-90,isosceles triangle stretches=true,inner sep=0.8pt,minimum width=0.25cm,minimum height=2mm]
\tikzstyle{vtriang}=[fill=yellow,draw=black,shape=isosceles triangle,shape border rotate=180,isosceles triangle stretches=true,inner sep=0.8pt,minimum width=0.25cm,minimum height=2mm]
\tikzstyle{vrt}=[fill=yellow,draw=black,shape=isosceles triangle,shape border rotate=0,isosceles triangle stretches=true,inner sep=0.8pt,minimum width=0.25cm,minimum height=2mm]
\tikzstyle{H box}=[rectangle,fill=yellow,draw=black,xscale=0.8,yscale=0.8, inner sep=0.6pt]
\tikzstyle{gbox}=[rectangle,fill=green,draw=black,xscale=1.0,yscale=1.0, inner sep=1.pt]
\tikzstyle{rbox}=[rectangle,fill=red,draw=black,xscale=1.0,yscale=1.0, inner sep=1.pt]
\tikzstyle{triangle}=[fill=yellow,draw=black,shape=isosceles triangle,shape border rotate=90,isosceles triangle stretches=true,inner sep=0.8pt,minimum width=0.25cm,minimum height=2mm]

\tikzstyle{bn}=[circle,fill=black,draw=black,scale=.8]
\tikzstyle{wn}=[circle,fill=white,draw=black,scale=.6]
\tikzstyle{dn}=[circle,fill=none,draw=gray]
\tikzstyle{bspider}=[fill=black,draw=black,scale=1,shape=isosceles triangle,shape border rotate=-90,isosceles triangle stretches=true,inner sep=1pt,minimum width=0.4cm,minimum height=3mm]
\tikzstyle{dbspider}=[fill=black,draw=black,scale=1,shape=isosceles triangle,shape border rotate=90,isosceles triangle stretches=true,inner sep=1pt,minimum width=0.4cm,minimum height=3mm]
\tikzstyle{L}=[rectangle,shape=rectangle,fill=green,draw=black]

\tikzstyle{Z dot}=[inner sep=0mm, minimum size=2mm, shape=circle, draw=black, fill={rgb,255: red,221; green,255; blue,221}]
\tikzstyle{Z phase dot}=[minimum size=5mm, font={\footnotesize\boldmath}, shape=rectangle, rounded corners=2mm, inner sep=0.2mm, outer sep=-2mm, scale=0.8, draw=black, fill={rgb,255: red,221; green,255; blue,221}]
\tikzstyle{X dot}=[Z dot, shape=circle, draw=black, fill={rgb,255: red,255; green,136; blue,136}]
\tikzstyle{X phase dot}=[Z phase dot, fill={rgb,255: red,255; green,136; blue,136}, font={\footnotesize\boldmath}]
\tikzstyle{hadamard edge}=[-, dashed, dash pattern=on 2pt off 0.5pt, thick, draw={rgb,255: red,68; green,136; blue,255}]

\tikzstyle{black dot}=[inner sep=0.7mm,minimum width=0pt,minimum height=0pt,fill=black,draw=black,shape=circle]

\tikzstyle{dot}=[black dot]
\tikzstyle{smalldot}=[inner sep=0.4mm,minimum width=0pt,minimum height=0pt,fill=black,draw=black,shape=circle]
\tikzstyle{white dot}=[dot,fill=white]
\tikzstyle{antipode}=[white dot,inner sep=0.3mm,font=\footnotesize]
\tikzstyle{smallwhitedot}=[smalldot,fill=white]
\tikzstyle{alt white dot}=[white dot,label={[xshift=3.07mm,yshift=-0.05mm,font=\footnotesize]left:$*$}]
\tikzstyle{gray dot}=[dot,fill=gray!40!white]
\tikzstyle{smallgraydot}=[smalldot,fill=gray!40!white]
\tikzstyle{box vertex}=[draw=black,rectangle]
\tikzstyle{small box}=[box vertex,fill=white]
\tikzstyle{whitebg}=[fill=white,inner sep=2pt]
\tikzstyle{graph state vertex}=[sg vertex,fill=black]

\tikzstyle{wide copoint}=[fill=white,draw=black,shape=isosceles triangle,shape border rotate=90,isosceles triangle stretches=true,inner sep=1pt,minimum width=1.5cm,minimum height=5mm]
\tikzstyle{wide point}=[fill=white,draw=black,shape=isosceles triangle,shape border rotate=-90,isosceles triangle stretches=true,inner sep=1pt,minimum width=1.5cm,minimum height=4mm]
\tikzstyle{very wide copoint}=[fill=white,draw=black,shape=isosceles triangle,shape border rotate=-90,isosceles triangle stretches=true,inner sep=1pt,minimum width=2.5cm,minimum height=4mm]
\tikzstyle{very wide empty copoint}=[draw=black,shape=isosceles triangle,shape border rotate=-90,isosceles triangle stretches=true,inner sep=1pt,minimum width=2.5cm,minimum height=4mm]
\tikzstyle{symm}=[ultra thick,shorten <=-1mm,shorten >=-1mm]

\tikzstyle{square box}=[rectangle,fill=white,draw=black,minimum height=5mm,minimum width=5mm,font=\small]
\tikzstyle{square gray box}=[rectangle,fill=gray!30,draw=black,minimum height=6mm,minimum width=6mm]
\tikzstyle{copoint}=[regular polygon,regular polygon sides=3,draw=black,scale=0.75,inner sep=-0.5pt,minimum width=7mm,fill=white]
\tikzstyle{point}=[regular polygon,regular polygon sides=3,draw=black,scale=0.75,inner sep=-0.5pt,minimum width=7mm,fill=white,regular polygon rotate=180]
\tikzstyle{gray point}=[point,fill=gray!40!white]
\tikzstyle{gray copoint}=[copoint,fill=gray!40!white]

\newcommand{\edgearrow}{{\arrow[black]{>}}}
\newcommand{\edgetick}{{\arrow[black,scale=0.7,very thick]{|}}}

\tikzstyle{diredge}=[->]
\tikzstyle{rdiredge}=[<-]
\tikzstyle{medium diredge}=[->]

\tikzstyle{short diredge}=[->]
\tikzstyle{halfedge}=[-)]
\tikzstyle{other halfedge}=[(-]
\tikzstyle{freeedge}=[(-)]
\tikzstyle{white edge}=[line width=5pt,white]
\tikzstyle{tick}=[postaction=decorate,decoration={markings, mark=at position 0.5 with \edgetick}]
\tikzstyle{small map edge}=[|-latex, gray!60!blue, shorten <=0.9mm, shorten >=0.5mm]
\tikzstyle{thick dashed edge}=[very thick,dashed,gray!40]
\tikzstyle{map edge}=[|-latex,very thick, gray!40, shorten <=1mm, shorten >=0.5mm]
\tikzstyle{tickedge}=[postaction=decorate,
  decoration={markings, mark=at position 0.5 with \edgetick}]
\tikzstyle{dirtickedge}=[postaction=decorate,
  decoration={markings, mark=at position 0.5 with \edgetick},
  decoration={markings, mark=at position 0.85 with \edgearrow}]
\tikzstyle{dirdoubletickedge}=[postaction=decorate,
  decoration={markings, mark=at position 0.4 with \edgetick},
  decoration={markings, mark=at position 0.6 with \edgetick},
  decoration={markings, mark=at position 0.85 with \edgearrow}]

\makeatletter
\newcommand{\boxshape}[3]{%
\pgfdeclareshape{#1}{
\inheritsavedanchors[from=rectangle] 
\inheritanchorborder[from=rectangle]
\inheritanchor[from=rectangle]{center}
\inheritanchor[from=rectangle]{north}
\inheritanchor[from=rectangle]{south}
\inheritanchor[from=rectangle]{west}
\inheritanchor[from=rectangle]{east}
\backgroundpath{
\southwest \pgf@xa=\pgf@x \pgf@ya=\pgf@y
\northeast \pgf@xb=\pgf@x \pgf@yb=\pgf@y

\@tempdima=#2
\@tempdimb=#3

\pgfpathmoveto{\pgfpoint{\pgf@xa - 5pt + \@tempdima}{\pgf@ya}}
\pgfpathlineto{\pgfpoint{\pgf@xa - 5pt - \@tempdima}{\pgf@yb}}
\pgfpathlineto{\pgfpoint{\pgf@xb + 5pt + \@tempdimb}{\pgf@yb}}
\pgfpathlineto{\pgfpoint{\pgf@xb + 5pt - \@tempdimb}{\pgf@ya}}
\pgfpathlineto{\pgfpoint{\pgf@xa - 5pt + \@tempdima}{\pgf@ya}}
\pgfpathclose
}
}}

\boxshape{NEbox}{0pt}{8pt}
\boxshape{SEbox}{0pt}{-8pt}
\boxshape{NWbox}{8pt}{0pt}
\boxshape{SWbox}{-8pt}{0pt}
\makeatother

\tikzstyle{map}=[draw,shape=NEbox,inner sep=7pt]
\tikzstyle{mapdag}=[draw,shape=SEbox,inner sep=7pt]
\tikzstyle{maptrans}=[draw,shape=SWbox,inner sep=7pt]
\tikzstyle{mapconj}=[draw,shape=NWbox,inner sep=7pt]

\tikzstyle{probs}=[shape=semicircle,fill=gray!40!white,draw=black,shape border rotate=180,minimum width=1.2cm]

\tikzstyle{arrs}=[-latex,font=\small,auto]
\tikzstyle{arrow plain}=[arrs]
\tikzstyle{arrow dashed}=[dashed,arrs]
\tikzstyle{arrow bold}=[very thick,arrs]
\tikzstyle{arrow hide}=[draw=white!0,-]
\tikzstyle{arrow reverse}=[latex-]
\tikzstyle{cdnode}=[]

\tikzstyle{gn}=[dot,fill=green,minimum width=0.25cm,inner sep=0pt]
\tikzstyle{rn}=[dot,fill=red,inner sep=0pt,minimum width=0.25cm]

\tikzstyle{rc}=[dot,thick,fill=white,draw = red,minimum width=0.3cm,inner sep=0pt]
\tikzstyle{gc}=[dot,thick,fill=white,draw= green,inner sep=0pt,minimum width=0.3cm]
\tikzstyle{bc}=[dot,thick,fill=white,draw= blue,minimum width=0.3cm]

\tikzstyle{label}=[circle,fill=white,minimum width=0.3cm]

\tikzstyle{clocklabel}=[dot,fill=yellow,draw=black,font=\tiny,inner sep=0.75pt]

\tikzstyle{rsn}=[circle split,draw,fill=red,font=\tiny,inner sep=0.75pt]
\tikzstyle{gsn}=[circle split,draw,fill=green,font=\tiny,inner sep=0.75pt]
\tikzstyle{bsn}=[circle split,draw,fill=blue,font=\tiny,inner sep=0.75pt]

\tikzstyle{rsc}=[circle split,thick,draw= red,draw,fill=white,font=\tiny,inner sep=0.75pt]
\tikzstyle{gsc}=[circle split,thick,draw= green,draw,fill=white,font=\tiny,inner sep=0.75pt]
\tikzstyle{bsc}=[circle split,thick,draw= blue,draw,fill=white,font=\tiny,inner sep=0.75pt]

\tikzstyle{cnot}=[fill=white,shape=circle,inner sep=-1.4pt]
\tikzstyle{wire label}=[font=\tiny, auto]

\tikzstyle{square box2}=[rectangle,fill=white,draw=black,minimum height=5mm,minimum width=10mm,font=\small]

\tikzstyle{scalar}=[rectangle,shape=diamond,fill=white,draw=black, inner sep=0.8pt]

\newcommand{\bra}[1]{\ensuremath{\left\langle #1 \right|}}
\newcommand{\ket}[1]{\ensuremath{\left|  #1 \right\rangle}}

\tikzstyle{cdiag}=[matrix of math nodes, row sep=3em, column sep=3em, text height=1.5ex, text depth=0.25ex,inner sep=0.5em]
\tikzstyle{arrow above}=[transform canvas={yshift=0.5ex}]
\tikzstyle{arrow below}=[transform canvas={yshift=-0.5ex}]

\usepackage{placeins}
\usepackage{mathtools}
\usepackage{graphicx}
\usepackage{pgfplots}
\pgfplotsset{compat=1.6}
\usepackage{stmaryrd}
\usepackage[normalem]{ulem} 
\usepackage{soul} 

\Title{A Compositional Model of Consciousness based on Consciousness-Only}

\Author{Camilo Miguel Signorelli $^{1,2,*}$\orcidA{0000-0002-2110-7646}, Quanlong Wang $^{3,*}$, Ilyas Khan$^{3,4,*}$}
\AuthorNames{Camilo Miguel Signorelli and Quanlong Wang and Ilyas Khan}

\address{%
$^{1}$ \quad Department of Computer Science, University of Oxford\\
$^{2}$ \quad Cognitive Neuroimaging Unit, INSERM U992, NeuroSpin\\
$^{3}$ \quad Cambridge Quantum Computing Ltd\\
$^{4}$ \quad St Edmund’s College, University of Cambridge
}
\corres{Email: cam.signorelli@cs.ox.ac.uk; harny.wang@cambridgequantum.com; ilyas@cambridgequantum.com}

\abstract{Scientific studies of consciousness rely on objects whose existence is assumed to be independent of any consciousness.  On the contrary, we assume  consciousness to be fundamental, and that one of the main features of consciousness is characterized as being other-dependent. We set up a framework which naturally subsumes this feature by defining a compact closed category where morphisms represent conscious processes. These morphisms are a composition of a set of generators, each being specified by their relations with other generators, and therefore co-dependent. The framework is general enough and fits well into a compositional model of consciousness. Interestingly, we also show how our proposal may become a  step towards  avoiding the hard problem of consciousness, and thereby address the combination problem of conscious experiences.}


\begin{document}


\section{Introduction}

Despite scientific advances in understanding the objective neural correlates of consciousness \cite{Seth2018}, science has so far failed in recovering subjective features from objective and measurable correlates of consciousness. One example is the unity of consciousness. According to the phenomenology of consciousness  \cite{Searle2000,Bayne2012}, one of the most salient features of conscious experience is its unity: "any set of conscious states of a subject at a time is unified...by being aspects of a single encompassing state of consciousness" \cite{Bayne2012}. If someone experiences colour and noise, the experience of colour is not followed by the experience of noise separately, even though it might be sequential, but both are experienced together as different aspects/content of one single conscious experience. Current models postpone the explanation of that unity, assuming there will be further developments \cite{Crick1998}. In the meantime, they reduce conscious experience to neural events.

In this article, we present an alternative approach: consciousness as a fundamental process of nature. Our approach takes inspiration from the Yogacara school \cite{Lusthaus2002, Makeham2014}, conscious agents model \cite{Fields2018} and phenomenology \cite{Thompson2007, Varela1996}. In our framework, subjectivity, a key feature of consciousness is characterised as \textbf{other-dependent} or \textbf{co-dependent}, i.e. the nature of existence arising from causes and conditions that are interdependent between each other. Without falling into idealism or dualism, we propose that consciousness should be treated as a primary process.

To model the co-dependent nature, we propose a compositional model for consciousness. This model is based on symmetric monoidal categories (Section \ref{secc:Pre}), a.k.a Process Theory \cite{Coecke2013, Coecke2016}. At the core of process theory lies the principle of compositionality (Section \ref{secc:PC}). Compositionality defines the whole as compositions of the parts. These parts, however, are not trivial decompositions, they contain in themselves the very properties that define the whole (in our case, conscious processes compound other conscious processes). Parts and the whole are therefore defined together, they co-depend. Compositionality is thus a middle ground between reductionism and holism \cite{Signorelli2020}. This makes process theory and our compositional framework suitable for investigating the irreducible structural properties of conscious experience \cite{Prentner2019}.

Finally, our framework  intends to mathematize a few aspects of the phenomenology of conscious experience (Section \ref{processtc}) and target its major questions \cite{Yoshimi2007,Tsuchiya2020}. For instance, the unity of consciousness naturally arises as result of composition, and the combination of fundamental experiences is discussed in light of our framework (Section \ref{secc:Unity}). This new perspective of scientific models of consciousness invokes pure mathematical entities, avoiding any ontological claim of their physical substrates (Section \ref{secc:Con}).

\section{Category Theory and Process Theory}
\label{secc:Pre} 

In this section, we briefly introduce the basic notions of Category theory \cite{Awodey2006,Maclane1965}, process theory \cite{Coecke2016} and graphical calculus \cite{Coecke2011b}.

\subsection{Preliminaries}
\section*{Category}

A category  $\mathfrak{C}$  consists of:
\begin{itemize}
\item a class of objects $ob(\mathfrak{C})$;
\item for each pair of objects $A, B$, a set  $\mathfrak{C}(A, B)$ of morphisms from $A$ to $B$;
\item for each triple of objects $A, B, C$, a  composition map  
\[
\begin{array}{ccc} 
\mathfrak{C}(B, C) \times \mathfrak{C}(A, B)& \longrightarrow & \mathfrak{C}(A, C)\\
(g, f) & \mapsto & g\circ f;
\end{array}
\]
\item for each object $A$, an identity morphism $1_A \in \mathfrak{C}(A, A)$,
\end{itemize}
satisfying the following axioms:
\begin{itemize}
\item associativity:  for any $f\in  \mathfrak{C}(A, B),  g\in  \mathfrak{C}(B, C), h\in  \mathfrak{C}(C, D)$,  there holds $(h\circ g)\circ f=h\circ (g\circ f)$;
\item identity law: for any $f\in  \mathfrak{C}(A, B),  1_B\circ f=f=f\circ 1_A$.
\end{itemize}
A morphism $f\in  \mathfrak{C}(A, B)$ is an  \textit{  isomorphism} if there exists a  morphism $g\in  \mathfrak{C}(B, A)$ such that $g\circ f=1_A$ and $f\circ g=1_B$.  A \textit{product category} $\mathfrak{A} \times \mathfrak{B}$ can be defined componentwise by two categories $\mathfrak{A}$ and $\mathfrak{B}$.

\section*{Functor}

Given categories $\mathfrak{C}$ and $\mathfrak{D}$, a functor  $F: \mathfrak{C} \longrightarrow \mathfrak{D}$  consists of:
\begin{itemize}
\item a  mapping  
\[
\begin{array}{ccc} 
ob(\mathfrak{C})& \longrightarrow & ob(\mathfrak{D})\\
A & \mapsto & F(A);
\end{array}
\]

\item for each pair of objects $A, B$ of $\mathfrak{C}$,  a map
\[
\begin{array}{ccc} 
\mathfrak{C}(A, B) & \longrightarrow &\mathfrak{D}(F(A), F(B))\\
f & \mapsto & F(f),
\end{array}
\]

\end{itemize}
satisfying the following axioms:
\begin{itemize}
\item preserving composition:  for any morphisms $f\in  \mathfrak{C}(A, B),  g\in  \mathfrak{C}(B, C)$,  there holds $F(g\circ f)=F(g)\circ F(f)$;
\item preserving identity: for any object $A$ of $\mathfrak{C}$,  $ F(1_A)=1_{F(A)}$.

\end{itemize}

A functor  $F: \mathfrak{C} \longrightarrow \mathfrak{D}$ is  \textit{ faithful (full)} if for each pair  of objects $A, B$ of $\mathfrak{C}$, the map
\[
\begin{array}{ccc} 
\mathfrak{C}(A, B) & \longrightarrow &\mathfrak{D}(F(A), F(B))\\
f & \mapsto & F(f)
\end{array}
\]
is injective (surjective). 

A bifunctor (also called binary functor) is just a functor whose domain is the  product of two categories.

\section*{Natural transformation}

Let $F, G: \mathfrak{C}\longrightarrow \mathfrak{D} $ be two functors. A natural transformation $\tau : F \rightarrow G$ is a family $(\tau_A:F(A)\longrightarrow G(A))_{A\in \mathfrak{C}} $ of morphisms in $ \mathfrak{D}$ such that the following square commutes: 

\begin{center}
\begin{tikzpicture}
\node (1) at  ( -0.5,4) [] {$F(A)$};
\node  at  ( 0.5,4.2) [] {$\tau_A$};
\node  at  ( -1,3.3) [] {$F(f)$};
\node (2) at  ( 1.5,4) [] {$G(A)$};
\node (3) at ( -0.5,2.5) [] {$F(B)$};
\node  at  ( 0.5,2.72) [] {$\tau_B$};
\node  at  ( 2,3.3) [] {$G(f)$};
\node (4) at ( 1.5,2.5) [] {$G(B)$};

\draw [->] (1) -- (2) {} ;
\draw [->] (3) -- (4) {} ;
\draw [->] (1) -- (3) {} ;
\draw [->] (2) -- (4) {} ;
\end{tikzpicture}
\end{center}
\noindent

for all morphisms $f\in \mathfrak{C}(A,B) $. A natural isomorphism is a natural transformation where each of the $\tau_A$ is an isomorphism. 

\section*{Strict monoidal category}

A strict monoidal category consists of: 

\begin{itemize}

\item a category $\mathfrak{C}$; 

\item a unit object $I\in ob(\mathfrak{C})$;

\item  a bifunctor $- \otimes - : \mathfrak{C} \times \mathfrak{C} \longrightarrow \mathfrak{C} $, 

\end{itemize}
satisfying 
\begin{itemize}
\item associativity:  for each triple of objects $A, B, C$ of $\mathfrak{C}$, $A\otimes (B \otimes C)=(A\otimes B) \otimes C$; for each triple of morphisms $f, g, h$ of $\mathfrak{C}$, $f\otimes (g \otimes h)=(f\otimes g) \otimes h$; 

\item unit law: for each object $A$ of $\mathfrak{C}$,  $A\otimes I= A=I\otimes A$; for each morphism $f$ of $\mathfrak{C}$,  $f\otimes 1_I= f=1_I\otimes f$.
\end{itemize}

\section*{Strict symmetric monoidal category}

A strict monoidal category $\mathfrak{C}$ is symmetric if it is equipped with a natural isomorphism

\begin{center}
$\sigma_{A,B} : A \otimes B \rightarrow B \otimes A$

\end{center}

\noindent  for all objects $A, B, C$ of $\mathfrak{C}$  satisfying:
 $$\sigma_{B,A} \circ \sigma_{A,B} =1_{A \otimes B},  ~~\sigma_{A,I}=1_A, ~~ (1_B \otimes \sigma_{A,C}) \circ  (\sigma_{A,B} \otimes 1_C)  =\sigma_{A, B\otimes C}.$$

\section*{Strict monoidal functor}
Given two strict monoidal categories $\mathfrak{C}$ and $\mathfrak{D}$, a strict monoidal functor  $F: \mathfrak{C} \longrightarrow \mathfrak{D}$  is a functor  $F: \mathfrak{C} \longrightarrow \mathfrak{D}$ such that 
$F(A)\otimes F(B)=F(A\otimes B), F(f)\otimes F(g)=F(f\otimes g), F(I_{\mathfrak{C}})=I_{\mathfrak{D}}$, for any objects $A, B$ of $\mathfrak{C}$, and any morphisms $f\in\mathfrak{C}(A, A_1), g\in\mathfrak{C}(B, B_1)$.

A strict symmetric monoidal functor $F$ is a strict monoidal functor that preserves symmetrical structures, i.e., $ F(\sigma_{A,B})= \sigma_{F(A),F(B)}$. The definition of a general (non-strict) symmetric monoidal functor can be found in \cite{Maclane1965}.

\section*{Strict compact closed category }

A strict compact closed category  is a strict symmetric monoidal category $\mathfrak{C}$ such that for each object $A$ of $\mathfrak{C}$, there exists a object $A^{*}$ and two morphisms
 $$\epsilon_{A} : A \otimes A^{*} \rightarrow I,     ~~            \eta_{A} : I \rightarrow  A^{*} \otimes A$$
 satisfying:

$$  (\epsilon_{A}   \otimes 1_A ) \circ (1_A \otimes \eta_A) =1_A, ~~         (1_A^{*} \otimes  \epsilon_{A}  ) \circ (\eta_A  \otimes 1_A^{*}) =1_A^{*}.$$
A strict compact closed category  is called self-dual if $A=A^{*}$ for each object $A$ \cite{Coecke2017a}.

\subsection{Process Theory}
\label{sec:process}
Process theory is an abstract framework of how things happen, be they mental or physical and regardless of their nature. Process theory describes how processes are composed. It has been widely used in various research fields such as the foundations of physical theories \cite{Coecke2011},  quantum theory \cite{Abramsky2004,Coecke2017a}, causal models \cite{Kissinger2017a, Pinzani2019}, relativity \cite{Kissinger2017} and interestingly also natural language \cite{Coecke2010} and cognition \cite{Bolt2017,Signorelli2020a}. In common with all theories,  process theory has its own assumptions, albeit with the advantage that its major feature is that it contains minimal assumptions. 

In process theory, we first assume an event occurs, i.e., a change from something typed as $A$ to something typed as $B$. This is called a process and denoted as a box:

$$%
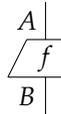
\beginpgfgraphicnamed{TikZit//singleprocess2}
\InputIfFileExists{TikZit//singleprocess2.tikz}{}{\input{./figures/TikZit//singleprocess2.tikz}}
\endpgfgraphicnamed$$

Second, we assume somethings happen sequentially, such as a process $g$ happens before another process $f$:  

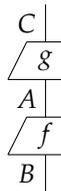
  $$%
\beginpgfgraphicnamed{TikZit//sequentialproc2}
\InputIfFileExists{TikZit//sequentialproc2.tikz}{}{\input{./figures/TikZit//sequentialproc2.tikz}}
\endpgfgraphicnamed$$
  
 $f$ happens after  $g$ can be seen as a single process from type $C$ to type $B$, which is denoted by $f\circ g: C\rightarrow B$. This is called \textbf{sequential composition}. As such, three things happening in sequence is seen as one process without any ambiguity, i.e., the sequential composition of processes is associative: $(f\circ g) \circ h=f\circ (g \circ h)$. We also assume that for each type $A$, there exists a process called the identity $1_A$, which does nothing at all to $A$. This is depicted as a straight line:
 
  $$ %
\beginpgfgraphicnamed{TikZit//identitya}
\begin{tikzpicture}
	\begin{pgfonlayer}{nodelayer}
		\node [style=none] (0) at (0.25, -0.5) {$A$};
		\node [style=none] (1) at (0, 0.5) {};
		\node [style=none] (2) at (0, -0.5) {};
	\end{pgfonlayer}
	\begin{pgfonlayer}{edgelayer}
		\draw (2.center) to (1.center);
	\end{pgfonlayer}
\end{tikzpicture}}
\endpgfgraphicnamed$$
  
As a consequence, given a process $f: A\rightarrow B$, we have $1_B\circ f = f = f\circ 1_A$ . 
  
Third, we assume that there should be different \textit{processes} happening simultaneously. Two processes $f$ and $g$ that happen simultaneously are described as:
$$  %
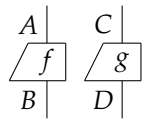
\beginpgfgraphicnamed{TikZit//parallelpro2}
\InputIfFileExists{TikZit//parallelpro2.tikz}{}{\input{./figures/TikZit//parallelpro2.tikz}}
\endpgfgraphicnamed$$

If we view two types, say $A$ and $C$, as a single type which we denote as $A\otimes C$, then the simultaneous processes $f$ and $g$ is a single process from type $A\otimes C$ to type $B\otimes D$, that we denote as $f\otimes g: A\otimes C \rightarrow B\otimes D$.  We call this a \textbf{parallel composition} of processes. 

The above depiction of $f\otimes g$ is asymmetric: $f$ on the left while $g$ on the right. This is due to the limitation of a planar drawing. If we want two processes that occur simultaneously placed in a symmetric way, it would mean that if we swap their positions, they should be essentially the same where all the types should match. This can be realised by adding a \textbf{swap} process:  $$%
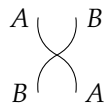
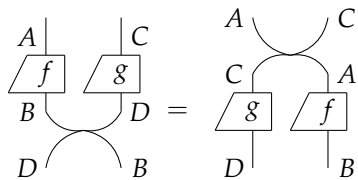
\beginpgfgraphicnamed{TikZit//swap2v}
\InputIfFileExists{TikZit//swap2v.tikz}{}{\input{./figures/TikZit//swap2v.tikz}}
\endpgfgraphicnamed$$ such that
  $$%
\beginpgfgraphicnamed{TikZit//swap2process2}
\InputIfFileExists{TikZit//swap2process2.tikz}{}{\input{./figures/TikZit//swap2process2.tikz}}
\endpgfgraphicnamed $$

With these basic assumptions, processes can be organised into what is called a \textbf{process theory} in the framework of a strict symmetric monoidal category (SMC). A much more detailed description of process theory can be found in \cite{Coecke2017a}.

To relate sequential and paralell composition in a simple way, one can add the compact structure to the process theory by using \textbf{caps} %
\beginpgfgraphicnamed{TikZit//capdit}
\begin{tikzpicture}
	\begin{pgfonlayer}{nodelayer}
		\node [style=none] (0) at (0.5, -0.25) {};
		\node [style=none] (1) at (-0.5, -0.25) {};
		\node [style=none] (2) at (-0.75, -0.25) {$s$};
		\node [style=none] (3) at (0.75, -0.25) {$s$};
	\end{pgfonlayer}
	\begin{pgfonlayer}{edgelayer}
		\draw [in=90, out=90, looseness=1.75] (1.center) to (0.center);
	\end{pgfonlayer}
\end{tikzpicture}}
\endpgfgraphicnamed and \textbf{cups} %
\beginpgfgraphicnamed{TikZit//cupdit}
\begin{tikzpicture}
	\begin{pgfonlayer}{nodelayer}
		\node [style=none] (0) at (0.75, 0.25) {$s$};
		\node [style=none] (1) at (-0.75, 0.25) {$s$};
		\node [style=none] (2) at (-0.5, 0.25) {};
		\node [style=none] (3) at (0.5, 0.25) {};
	\end{pgfonlayer}
	\begin{pgfonlayer}{edgelayer}
		\draw [in=-90, out=-90, looseness=1.75] (2.center) to (3.center);
	\end{pgfonlayer}
\end{tikzpicture}}
\endpgfgraphicnamed, so that:

$$  %
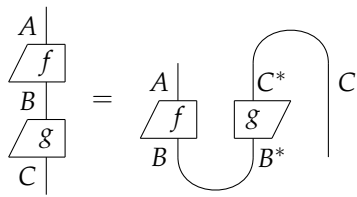
\beginpgfgraphicnamed{TikZit//timespace2}
\InputIfFileExists{TikZit//timespace2.tikz}{}{\input{./figures/TikZit//timespace2.tikz}}
\endpgfgraphicnamed $$

Mathematically speaking, we now have a compact closed category.  

As introduced above, process theory focuses on the processes instead of the objects/types, providing a philosophical advantage: process theories emphasise transformations, avoiding any ontological claim or \textit{substance-like} description of invariant properties of those types, like mass and charge. 

\subsection{Fine-grained Version of Process Theory}
\label{secc:Fine}
In general process theory, most of the boxes (processes) are unspecified in the sense that what is inside a box is unknown, whereas we need to know more details about their interactions in some applications. In other words, we need a fine-grained version of process theory. The typical way to derive such a version is to generate all the processes by a set of basic processes called \textbf{generators}, while specifying those generators in terms of equations of processes composed of generators. Below, we illustrate this idea by a typical example called ZX-calculus. 

ZX-calculus is a process theory invented by Bob Coecke and Ross Duncan as a graphical language for a pair of complementary quantum processes (represented by two diagrams called green spider and red spider respectively) \cite{Coecke2011b}. All the processes in ZX-calculus are diagrams composed sequentially or in parallel, either of green spiders with phase parameters, red spiders with phase parameters,  straight lines, swaps, caps or cups. 

These generators satisfy a set of diagrammatic equations called \textbf{rewriting rules}: one can  rewrite each diagram into an equivalent one by replacing a part of the diagram which is on one side of an equation with the diagram on the other side of the equation. All the ZX diagrams modulo \footnote{Modulo means using an equivalent relation.} and the rewriting rules form a self-dual compact closed category \cite{Coecke2011b}. To guarantee that there are no conflicts in this rewriting system, ZX-calculus needs a property called \textbf{soundness}: there exists a \textbf{standard interpretation} from the category of ZX diagrams to the category of matrices, i.e., a  symmetric monoidal functor between them \cite{Coecke2011b}. More general, a sound rewriting system means that there are not internal contradictions, while \textbf{completeness} would mean that we can prove anything that is right about the phenomena in question with the chosen system. A sound and complete rewriting system defines a unique set of generators.

\section{Compositional approach for consciousness-only}\label{secc:PC}

In this section, we motivate and explain the concepts of consciousness as fundamental and also the structure for consciousness given by the Yogacara School.

 \subsection{Process Theory for consciousness}
 \label{secc:PZX}
 
In any attempt to model consciousness, we expect to fulfill at least three theoretical requirements. First, one would like a theory with a basic and minimum set of assumptions. Process theory seems to fit with that requirement. Symmetric monoidal categories start from a minimum and specific intuitive form to deal with compositions, sequential and parallel. In the rest of this work, we will assume that this minimal structure already convey part of the \textit{experience structure}. This assumption is partially justified in the fact that, despite a unified experience, we only experience things happening in sequence (one after the other, time) or in parallel (side by side, space). 

Second, one would expect those minimum assumptions to be explicit. In other words, we need to model the nature of consciousness from explicit, primitive and axiomatic principles. Process theory in particular, and category theory in general, provides us with an exceptionally well suited mathematics for such axiomatic purposes. Due to the minimality of those assumptions,  any extra structure to be added to a process theory will also have an explicit mathematical meaning.
 
Third, one would like to recover and describe important properties of consciousness from those basic and explicit axioms. Specifically, we would like to recover the unity of consciousness. In process theory, compositionality outlines any unity as a non-trivially composition of some basic processes \cite{Coecke2013, Coecke2016}. Unity is formed by sequential and parallel compositions of \textit{primary} processes. Due to this foundational aspect, compositionality may be a convenient way to target the unity of consciousness, by modelling the unity of experience inside a process theory (section \ref{secc:Unity}).

\subsection{Consciousness as Fundamental}
  \label{secc:CF}

Natural science has achieved great success in modern age, behind which there lies a basic assumption that everything is made of physical objects whose existence is independent of any consciousness. This assumption is so powerful that renders scientific theories not too complicated and allows science results to be tested by independent experiments. However, such objective existence can never be verified by conscious agents, since most  cognitive activity is done through consciousness \cite{Signorelli2018}, thus the assumption is totally suspended.

Objectivity relates to a perceived or unperceived object, while subjectivity to a perceiving subject. In a everyday understanding of the terms, the object is meant to exist independently of any subject to perceive it (ontology), and as such, objectivity is commonly associated with concepts like truth and reliability (epistemic) \cite{Mulder,Searle1998}, e.g. the visible wavelength of light range from $~700$nm to $~400$nm. Contrary, subjectivity seems always interdependent, it involves both perceived and perceiving aspects, making subjective properties dependent of others interactions (internal or external to the subject) and thereof co-dependent \cite{Signorelli2020}. We can further distinguish between epistemic subjectivity, i.e. claims not verifiable (e.g. the claim that red is more beautiful than green), and ontological subjectivity \cite{Searle1998}: subjective modes of existence, such as pain or "redness" experienced only by the subject.

In order to understand our point of departure, we first need to recognize that objectivity is an assumption of basic science. The assumption of objectivity as primitive or fundamental is deeply grounded in neuroscience, as well as other scientific fields \cite{Anderson1972, Mazzocchi2008, BeimGraben2016}. Taking that assumption, one would expect that the subjective aspects of the experience may naturally emerge from the interaction and combination of physical objective entities. For example, contemporary theories of consciousness tend to focus on the physical \textit{atomic} parts from which, for instance, the unity of experience would emerge as a whole. The parts are considered cells, neurons, brain regions, and the whole being the unified conscious experience. This is called building blocks models \cite{Searle2000} or reductionist approaches \cite{Mazzocchi2008}. This approach, however, leads to the hard problem of consciousness \cite{ThomasNagel1974, Searle2000, Chalmers1995}: since each neuron is composed of basic physical objects--atoms,  and they are radically different from consciousness in that they lack key features of the latter like self-awareness and unity  \cite{Searle2000,Bayne2012},  no matter how complicated the interaction of these physical parts could be, how can those key features of consciousness arise from them? At this point, we consider the reader is familiar enough with this problem, so that we can avoid any deeper introduction.

An alternative assumption is to treat conscious experience as primary, or fundamental process of nature. We assume that all primary objects are indeed conscious-dependent. Treating conscious experience as primary convey two possible interpretations: i) ontological, i.e. the nature/existence of consciousness is fundamental (substance), ii) epistemic, i.e. the nature of knowledge about the world is limited by our experience. In this line of thoughts "our knowledge is limited to the realm of our own subjective impressions, allowing us no knowledge of objective reality in and of itself" \cite{Mulder, Varela1996}. In this paper, we are neutral about what is the optimal interpretation. Independently, we emphasize that conscious experience is a primary \textit{process} of nature, a transformation. Being fundamental would also means that there is not further explanation. Therefore, physical objects would be the result of \textit{consciousness transforming} and everything considered, affirmed or denied, even the idea of objectivity, would occur to us only in consciousness.

Although the new assumption dissolves or evades the hard problem of consciousness, it comes with what we call the \textit{dual problem of consciousness}: the question about how the objective realm arise from subjective one. To deal with that problem and model conscious experience from the assumption of the primacy of consciousness, we take inspiration from the Eastern philosophy known as Yogacara.

 \subsection{Yogacara Philosophy}
 \label{secc:YPC}
 
The reason to choose the Yogacara philosophy and its phenomenology is mainly because it has an explicit description of a structure of eight types of consciousnesses and the relation between consciousness and the physical world. Moreover, the key feature of the Yogacara philosophy is consciousness-only which means there is nothing outside of all sentient beings' consciousnesses.  In modern words, consciousness-only would be better understood as a claim of awareness-only, or perception-only, much closer to modern phenomenology \cite{Lusthaus2002,LI2015,Kern1988}.
 
The Yogacara philosophy has a rich system of eight consciousnesses consisting of: the first seven consciousnesses---the five \textbf{sense-consciousnesses} (eye or visual, ear or auditory, nose or olfactory, tongue or gustatory, body or tactile consciousnesses), \textbf{mental consciousness} (the sixth consciousness),  \textbf{manas consciousness} (the seventh or thought-centre consciousness), and the eighth consciousness---\textbf{alaya consciousness} (storehouse  consciousness). These eight consciousnesses are not independent of each other: "... the Alaya consciousness and the first seven consciousnesses generate each in a steady process and are reciprocally cause and effect" \cite{Francis}. A clarifying metaphor is to think about the eighth consciousness as the ocean, while the other consciousness are different types of waves in its surface. Neither of them are separated of the others and all consciousness are essentially one.
 
 In this framework, the act of perception of the eighth consciousness (Alaya consciousness) is considered extremely subtle or difficult to perceive \cite{TatWei}. Alaya consciousness is thought to be the \textit{seed consciousness}, i.e. to contain on its own different \textbf{potentialities} that would engender other complex types of experiences \cite{TatWei,Lusthaus2002}. We will approach these potentialities only in relation with other seeds, leaving the \textit{types} on our process theory for Alaya consciousness unspecified \footnote{A future approach may define the internal structure of Alaya taking six features in formal analogy with the seed metaphor from \cite{TatWei}.}. This might be an economical strategy, since, although this structure is considered the same for all living beings, the input and outputs types for those processes might be species dependent, or even specific to each individual.  
 
 Moreover, each consciousness "manifests itself in two functional divisions (bhgas), namely, image and perception, i.e., the object perceived or \textbf{perceived division} and the perceiving faculty or \textbf{perceiving division} (nimittabhaga and darsanabhaga)" \cite{TatWei}. The perceived is related to the object and  the perceiving to the subject. In Husserlian phenomenology, this division is extrapolated to what is called Noema versus Noesis distinction \cite{Husserl1983}. The first division is mostly related to the sixth consciousness and the five perceptual consciousnesses, while the second one with the seventh manas consciousness.
  
 The phenomenon of the physical world and the body which we feel everyday comes from the perceived division of Alaya consciousness: "it transforms internally into seeds and the body provided with organs, and externally into the world receptacle. These things that are its transformations become its own object of perception (dlanzbana)"  \cite{Francis}. The receptacle-world and the Body as part of the perceived division of Alaya consciousness  should not be thought of as the physical world and  the physical body that we feel in our normal lives, but as being related in that the appearance of the latter is based on the existence of the former. As a consequence, the objectivity of the world comes from the same structure shared by different sentient beings in the perceived division of their Alaya consciousnesses. 
  
 In the rest of this work, we will focus on a simple model for the perceived and perceiving division of Alaya consciousness. In order to have a model of this structure, we highlight three key ideas: i) Alaya consciousness is very subtle and it is only shown before us as co-dependent or interdependent process, ii) Alaya consciousness is primary/fundamental, from which other consciousness and the physical world may arise, iii) Specifically, the physical world arise from the perceived division of Alaya consciousness.

\section{Compositional Model for Consciousness-Only}
\label{processtc} 

After the discussion in section \ref{secc:PC}, we now provide a compositional model of consciousness based on the Yogacara philosophy of consciousness-only and a few further assumptions.

\subsection{Process Theory for Alaya Consciousness}
\label{secc:MCG}

The first feature of Alaya consciousness is its co-dependence, which means each process of Alaya consciousness is dependent on other processes. The general process theory can not display the other-dependence feature because most of its processes are not specified (see section \ref{secc:Fine}). So we need a fine-grained version of process theory which has generators specified by explicit rewriting rules. We might also choose these generators for \textit{non-classical} systems. This choice is partially justified by recent models of psychology and cognition that seems to be quantum related \cite{Bruza2015,Cervantes2018}. Moreover, we also require that any parameter appeared in the theory is not a concrete number, according to the unspecification of the types we discussed above. 

Based on the requirements for a fine-grained process theory that are noted in previous sections, we introduce a formalism called qufinite $ZX_{\Delta}$-calculus, which is a generalisation of the normal ZX-calculus  \cite{Coecke2011b} regarding the following aspects: 1) a labelled triangle symbol is introduced as a new generator, that's why there is a $\Delta$ in the name of the generalised ZX-calculus, 2) all the qudit ZX-calculus (ZX-calculus for qudits-- quantum versions of d-ary digits) are unified in a single framework, 3) the parameters (phases) of normal ZX-calculus are generalised from complex numbers to elements of an arbitrary commutative semiring. 

We give the details below of the  qufinite $ZX_{\Delta}$-calculus: generators and rewriting rules. Throughout this section, $\mathbb N =\{0, 1, 2, \cdots\}$ is the set of natural numbers,  $2 \leq d  \in \mathbb N$, $\oplus$ is the modulo $d$ addition, $\mathcal{S}$ is an arbitrary commutative semiring \cite{Golan1999}.  All the diagrams are read from top to bottom as in previous sections.

\subsubsection{Generators of Qufinite $ZX_{\Delta}$-calculus}
\label{secc:Qufin1}

We give the generators of the qufinite $ZX_{\Delta}$-calculus  in Table \ref{qbzxgeneratordit}.

\begin{table}[!h]
\begin{center} 
\begin{tabular}{|r@{~}r@{~}c@{~}c|r@{~}r@{~}c@{~}c|}
\hline
&& & %
\beginpgfgraphicnamed{TikZit//generalgreenspiderqdit}
\InputIfFileExists{TikZit//generalgreenspiderqdit.tikz}{}{\input{./figures/TikZit//generalgreenspiderqdit.tikz}}
\endpgfgraphicnamed  & &&& %
\beginpgfgraphicnamed{TikZit//redspider0pd}
\InputIfFileExists{TikZit//redspider0pd.tikz}{}{\input{./figures/TikZit//redspider0pd.tikz}}
\endpgfgraphicnamed\\\hline
&& & %
\beginpgfgraphicnamed{TikZit//triangled}
\InputIfFileExists{TikZit//triangled.tikz}{}{\input{./figures/TikZit//triangled.tikz}}
\endpgfgraphicnamed  & &&& %
\beginpgfgraphicnamed{TikZit//redclassicd}
\InputIfFileExists{TikZit//redclassicd.tikz}{}{\input{./figures/TikZit//redclassicd.tikz}}
\endpgfgraphicnamed\\\hline
&& & %
\beginpgfgraphicnamed{TikZit//idqudit}
\InputIfFileExists{TikZit//idqudit.tikz}{}{\input{./figures/TikZit//idqudit.tikz}}
\endpgfgraphicnamed  & &&& %
\beginpgfgraphicnamed{TikZit//swapd}
\InputIfFileExists{TikZit//swapd.tikz}{}{\input{./figures/TikZit//swapd.tikz}}
\endpgfgraphicnamed\\\hline
&& & %
\beginpgfgraphicnamed{TikZit//capdit}
\InputIfFileExists{TikZit//capdit.tikz}{}{\input{./figures/TikZit//capdit.tikz}}
\endpgfgraphicnamed  & &&& %
\beginpgfgraphicnamed{TikZit//cupdit}
\InputIfFileExists{TikZit//cupdit.tikz}{}{\input{./figures/TikZit//cupdit.tikz}}
\endpgfgraphicnamed\\\hline
&& & %
\beginpgfgraphicnamed{TikZit//binderdit}
\InputIfFileExists{TikZit//binderdit.tikz}{}{\input{./figures/TikZit//binderdit.tikz}}
\endpgfgraphicnamed  & &&& %
\beginpgfgraphicnamed{TikZit//binderditflip}
\InputIfFileExists{TikZit//binderditflip.tikz}{}{\input{./figures/TikZit//binderditflip.tikz}}
\endpgfgraphicnamed\\\hline
\end{tabular} \caption{Generators of qufinite $ZX_{\Delta}$-calculus, where  $m,n\in \mathbb N; \protect\overrightarrow{\alpha_d}=(a_1,\cdots, a_{d-1}); a_i\in \mathcal{S}; i \in \{1,\cdots, d-1\}; j \in \{0, 1,\cdots, d-1\}; s, t \in \mathbb N \backslash\{0\}$. }\label{qbzxgeneratordit}
\end{center}
\end{table}

\FloatBarrier
\begin{Remark}
Each input or output of a generator is labeled by a positive integer. For simplicity, the first four generators have each of their inputs and outputs labelled by $d$, and we just give one label to a wire.
\end{Remark}

For simplicity, we use the following conventions: 
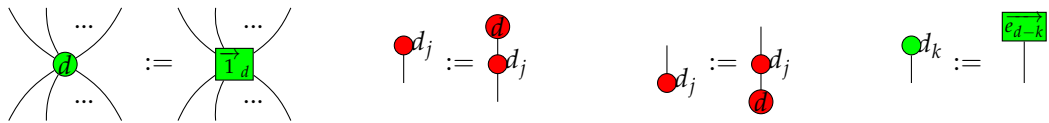
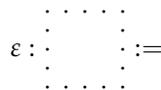
\[
\beginpgfgraphicnamed{TikZit//denotequfinite}
\InputIfFileExists{TikZit//denotequfinite.tikz}{}{\input{./figures/TikZit//denotequfinite.tikz}}
\endpgfgraphicnamed 
\]
and $$\varepsilon: %
\beginpgfgraphicnamed{TikZit//emptysquare}
\InputIfFileExists{TikZit//emptysquare.tikz}{}{\input{./figures/TikZit//emptysquare.tikz}}
\endpgfgraphicnamed:=$$
 
where $\overrightarrow{1}_d=\overbrace{(1,\cdots,1)}^{d-1};  j \in \{0, 1,\cdots, d-1\};  k\in \{1,\cdots, d-1\}; \overrightarrow{e_{d-k}}=\overbrace{(\underbrace{0,\cdots,1}_{d-k}, \cdots, 0)}^{d-1}$; $\varepsilon$ represents an empty diagram.

In terms of consciousness interpretation, each generator may be thought as a basic or primary conscious experience. Then, the set of generators becomes a minimal set of experiences. This selection is not unique, unless the group of generators is sound and complete, as we mentioned in section \ref{secc:Fine}.

For example, we may consider that basic conscious experiences involve many inputs and outputs types. This is realized by the many legs on %
\beginpgfgraphicnamed{TikZit//generalgreenspiderqdit}
\InputIfFileExists{TikZit//generalgreenspiderqdit.tikz}{}{\input{./figures/TikZit//generalgreenspiderqdit.tikz}}
\endpgfgraphicnamed  and %
\beginpgfgraphicnamed{TikZit//redspider0pd}
\InputIfFileExists{TikZit//redspider0pd.tikz}{}{\input{./figures/TikZit//redspider0pd.tikz}}
\endpgfgraphicnamed. The experience of adding types might be described by %
\beginpgfgraphicnamed{TikZit//triangled}
\begin{tikzpicture}
	\begin{pgfonlayer}{nodelayer}
		\node [style=none] (0) at (0, 0.5) {};
		\node [style=none] (1) at (0, -0.5) {};
		\node [style=triangle] (2) at (0, 0) {};
		\node [style=none] (3) at (-0.25, 0) {$d$};
	\end{pgfonlayer}
	\begin{pgfonlayer}{edgelayer}
		\draw (0.center) to (1.center);
	\end{pgfonlayer}
\end{tikzpicture}}
\endpgfgraphicnamed and perhaps primary perception realized by the process %
\beginpgfgraphicnamed{TikZit//redclassicd}
\begin{tikzpicture}
	\begin{pgfonlayer}{nodelayer}
		\node [style=none] (0) at (0, 0.75) {};
		\node [style=none] (1) at (0, -0.75) {};
		\node [style=rn] (2) at (0, 0) {};
		\node [style=none] (3) at (0.25, 0) {$d_j$};
	\end{pgfonlayer}
	\begin{pgfonlayer}{edgelayer}
		\draw (0.center) to (1.center);
	\end{pgfonlayer}
\end{tikzpicture}}
\endpgfgraphicnamed. The experience of \textit{inverting} types might correspond to caps and cups, combinations and segregation experiences may be represented by each of the rhomboids, respectively.

The concrete specification of these or other generators is an empirical task that we left for future works with trained phenomenologists. In our framework, some experiences that are considered basic, such as seeing red or hearing a monotone sound may indeed be the result of composition from our chosen set of generators or another set. This is due to the particular choice we have made. Moreover, one can also choose those monotone experiences (as far as one can give them explicit mathematical meaning) as part of another particular set of generators. The question of what is the unique set of phenomenal generatores is empirical and theoretical issue that may require attentive phenomenology and micro-phenomenological tools. In that case, the goal is to target the soundness and completeness of rewriting systems for conscious experience. Something far beyond the scope of our preliminary attempt.

\subsubsection{Rules of Qufinite $ZX_{\Delta}$-calculus }
\label{secc:Qufin2}

\begin{figure}[!h]
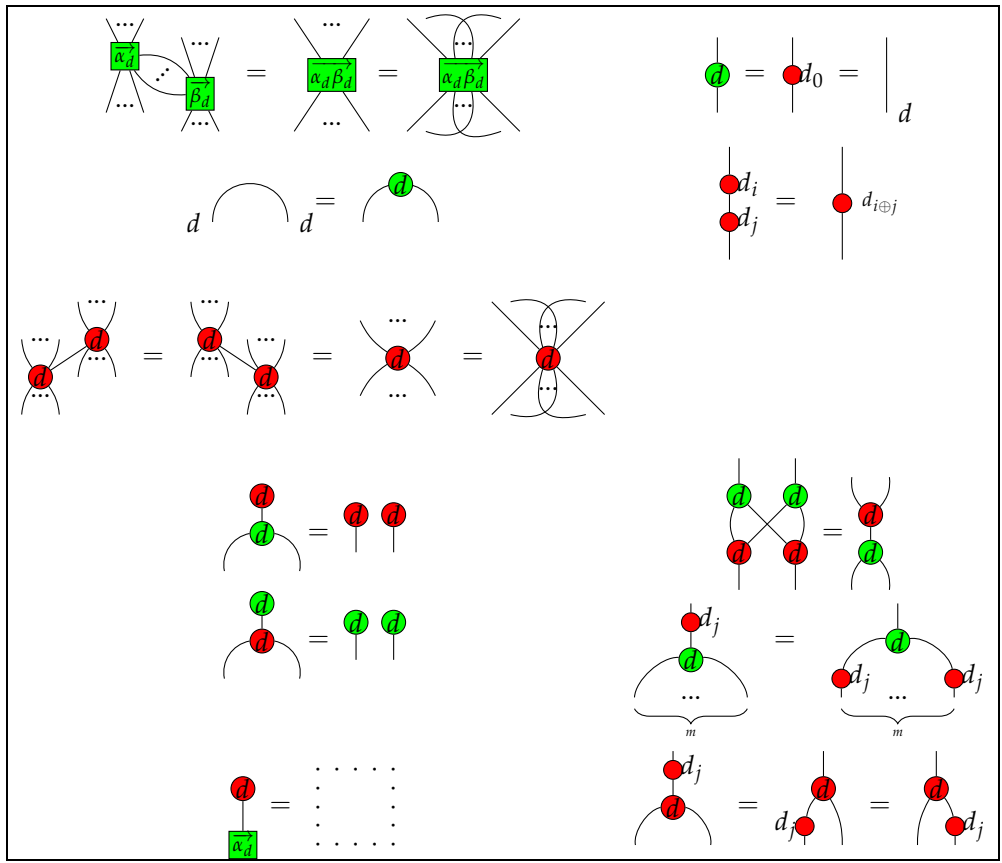

\begin{center} 
\[
\quad \qquad\begin{array}{|cc|}
\hline
\beginpgfgraphicnamed{TikZit//gengspiderfusedit}
\InputIfFileExists{TikZit//gengspiderfusedit.tikz}{}{\input{./figures/TikZit//gengspiderfusedit.tikz}}
\endpgfgraphicnamed &%
\beginpgfgraphicnamed{TikZit//s2qudit}
\InputIfFileExists{TikZit//s2qudit.tikz}{}{\input{./figures/TikZit//s2qudit.tikz}}
\endpgfgraphicnamed \\
\beginpgfgraphicnamed{TikZit//s3qudit}
\InputIfFileExists{TikZit//s3qudit.tikz}{}{\input{./figures/TikZit//s3qudit.tikz}}
\endpgfgraphicnamed & %
\beginpgfgraphicnamed{TikZit//rphaseaddqdit}
\InputIfFileExists{TikZit//rphaseaddqdit.tikz}{}{\input{./figures/TikZit//rphaseaddqdit.tikz}}
\endpgfgraphicnamed \\
&\\ 
\beginpgfgraphicnamed{TikZit//redspider0pfusedit}
\InputIfFileExists{TikZit//redspider0pfusedit.tikz}{}{\input{./figures/TikZit//redspider0pfusedit.tikz}}
\endpgfgraphicnamed&\\
&\\ 
\beginpgfgraphicnamed{TikZit//b1qudit}
\InputIfFileExists{TikZit//b1qudit.tikz}{}{\input{./figures/TikZit//b1qudit.tikz}}
\endpgfgraphicnamed & %
\beginpgfgraphicnamed{TikZit//b2qudit}
\InputIfFileExists{TikZit//b2qudit.tikz}{}{\input{./figures/TikZit//b2qudit.tikz}}
\endpgfgraphicnamed\\ 
%
\beginpgfgraphicnamed{TikZit//b3qudit}
\InputIfFileExists{TikZit//b3qudit.tikz}{}{\input{./figures/TikZit//b3qudit.tikz}}
\endpgfgraphicnamed  & %
\beginpgfgraphicnamed{TikZit//pimultiplecpdit}
\InputIfFileExists{TikZit//pimultiplecpdit.tikz}{}{\input{./figures/TikZit//pimultiplecpdit.tikz}}
\endpgfgraphicnamed\\
    %
\beginpgfgraphicnamed{TikZit//rdotaemptydit}
\InputIfFileExists{TikZit//rdotaemptydit.tikz}{}{\input{./figures/TikZit//rdotaemptydit.tikz}}
\endpgfgraphicnamed&%
\beginpgfgraphicnamed{TikZit//piasphasedit}
\InputIfFileExists{TikZit//piasphasedit.tikz}{}{\input{./figures/TikZit//piasphasedit.tikz}}
\endpgfgraphicnamed\\
  		  		\hline  
  		\end{array}\]      
  	\end{center}
  	\caption{Qufinite $ZX_{\Delta}$-calculus rules I, where $\protect\overrightarrow{\alpha_d}=(a_1,\cdots, a_{d-1}); \protect\overrightarrow{\beta_d}=(b_1,\cdots, b_{d-1}); \protect\overrightarrow{\alpha_d\beta_d}=(a_1b_1,\cdots, a_{d-1}b_{d-1}); a_k, b_k\in \mathcal{S}; k \in \{1,\cdots, d-1\}; j \in \{0, 1,\cdots, d-1\}; m  \in  \mathbb N.$}\label{qufiniterules1}
  \end{figure}

We provide rewriting rules for qufinite $ZX_{\Delta}$-calculus  in Figure \ref{qufiniterules1} and Figure \ref{qufiniterules2}. Even though we do not specify which generator corresponds to each \textit{basic} phenomenal experience, these rules specify the generators in light of what they do regarding each other. Concretely, here we focus on the idea that two or more generators define each other. For example, the green dot %
\beginpgfgraphicnamed{TikZit//singleddot}
\begin{tikzpicture}
	\begin{pgfonlayer}{nodelayer}
		\node [style=gn] (0) at (0, 0.25) {$d$};
		\node [style=none] (1) at (0, -0.25) {};
	\end{pgfonlayer}
	\begin{pgfonlayer}{edgelayer}
		\draw (0) to (1.center);
	\end{pgfonlayer}
\end{tikzpicture}}
\endpgfgraphicnamed  is specified by the rule %
\beginpgfgraphicnamed{TikZit//b3qudit}
\InputIfFileExists{TikZit//b3qudit.tikz}{}{\input{./figures/TikZit//b3qudit.tikz}}
\endpgfgraphicnamed  in the way that it is the only green spider which has no input and one output and can be copied by the red spider %
\beginpgfgraphicnamed{TikZit//singledrcopy}
\InputIfFileExists{TikZit//singledrcopy.tikz}{}{\input{./figures/TikZit//singledrcopy.tikz}}
\endpgfgraphicnamed. Moreover, the red spider %
\beginpgfgraphicnamed{TikZit//singledrcopy}
\InputIfFileExists{TikZit//singledrcopy.tikz}{}{\input{./figures/TikZit//singledrcopy.tikz}}
\endpgfgraphicnamed is also specified by the effects in the green dot %
\beginpgfgraphicnamed{TikZit//singleddot}
\begin{tikzpicture}
	\begin{pgfonlayer}{nodelayer}
		\node [style=gn] (0) at (0, 0.25) {$d$};
		\node [style=none] (1) at (0, -0.25) {};
	\end{pgfonlayer}
	\begin{pgfonlayer}{edgelayer}
		\draw (0) to (1.center);
	\end{pgfonlayer}
\end{tikzpicture}}
\endpgfgraphicnamed. It means that the experience of A, only makes sense if there is another experience B, from which one has a relationship with the other. For instance, the experience of the colour red only make sense if there is another experience of colour, for example, green and blue, otherwise there is not such colour experience at all, or at least, it is of very different nature (e.g. 
colour blindness). This is understood as a kind of contextual character of conscious phenomena, a particular aspect of experience. 

These rewriting rules become the axioms regarding a group of primary experiences, and further ways to define the generators in relation with their consciousness interpretation. In the example above, the red spider with many legs may convey the experience of \textit{copy} the experience from the green one. Again, specific phenomenal interpretations are left for future works, while we focus here on the introduction of the main concepts and the mathematical machinery.

 \FloatBarrier

 \begin{figure}[!h]
\begin{center} 
\[
\quad \qquad\begin{array}{|cc|}
\hline
\beginpgfgraphicnamed{TikZit//triangleocopydit}
\InputIfFileExists{TikZit//triangleocopydit.tikz}{}{\input{./figures/TikZit//triangleocopydit.tikz}}
\endpgfgraphicnamed &%
\beginpgfgraphicnamed{TikZit//trianglepicopydit}
\InputIfFileExists{TikZit//trianglepicopydit.tikz}{}{\input{./figures/TikZit//trianglepicopydit.tikz}}
\endpgfgraphicnamed \\
\beginpgfgraphicnamed{TikZit//sucdit}
\InputIfFileExists{TikZit//sucdit.tikz}{}{\input{./figures/TikZit//sucdit.tikz}}
\endpgfgraphicnamed & %
\beginpgfgraphicnamed{TikZit//zerotoreddit}
\InputIfFileExists{TikZit//zerotoreddit.tikz}{}{\input{./figures/TikZit//zerotoreddit.tikz}}
\endpgfgraphicnamed \\
&\\ 
\beginpgfgraphicnamed{TikZit//phasecopydit}
\InputIfFileExists{TikZit//phasecopydit.tikz}{}{\input{./figures/TikZit//phasecopydit.tikz}}
\endpgfgraphicnamed & %
\beginpgfgraphicnamed{TikZit//wsymetrydit}
\InputIfFileExists{TikZit//wsymetrydit.tikz}{}{\input{./figures/TikZit//wsymetrydit.tikz}}
\endpgfgraphicnamed\\ 
&\\ 
\beginpgfgraphicnamed{TikZit//associatedit}
\InputIfFileExists{TikZit//associatedit.tikz}{}{\input{./figures/TikZit//associatedit.tikz}}
\endpgfgraphicnamed  & %
\beginpgfgraphicnamed{TikZit//additiondit}
\InputIfFileExists{TikZit//additiondit.tikz}{}{\input{./figures/TikZit//additiondit.tikz}}
\endpgfgraphicnamed\\
&\\ 
\beginpgfgraphicnamed{TikZit//binderunitary1}
\InputIfFileExists{TikZit//binderunitary1.tikz}{}{\input{./figures/TikZit//binderunitary1.tikz}}
\endpgfgraphicnamed&%
\beginpgfgraphicnamed{TikZit//binderunitary2}
\InputIfFileExists{TikZit//binderunitary2.tikz}{}{\input{./figures/TikZit//binderunitary2.tikz}}
\endpgfgraphicnamed\\
    &\\ 
\beginpgfgraphicnamed{TikZit//binderassoc}
\InputIfFileExists{TikZit//binderassoc.tikz}{}{\input{./figures/TikZit//binderassoc.tikz}}
\endpgfgraphicnamed&%
\beginpgfgraphicnamed{TikZit//bindergspider}
\InputIfFileExists{TikZit//bindergspider.tikz}{}{\input{./figures/TikZit//bindergspider.tikz}}
\endpgfgraphicnamed\\
    &\\ 
  		  		\hline  
  		\end{array}\]      
  	\end{center}
  	\caption{Qufinite $ZX_{\Delta}$-calculus rules II,  where $\protect\overrightarrow{1}_d=\protect\overbrace{(1,\cdots,1)}^{d-1}; \protect\overrightarrow{0}_d=\protect\overbrace{(0,\cdots,0)}^{d-1};  \protect\overrightarrow{\alpha_d}=(a_1,\cdots, a_{d-1}); \protect\overrightarrow{\beta_d}=(b_1,\cdots, b_{d-1}); a_k, b_k\in \mathcal{S}; k \in \{1,\cdots, d-1\};  j \in \{ 1,\cdots, d-1\}; s, t, u \in \mathbb N \backslash\{0\}.$}\label{qufiniterules2}
  \end{figure}
 \FloatBarrier

Additionally, in order to form a compact closed category of diagrams, we also need the following structural rules:
 
 \begin{equation}\label{compactstructure}
\beginpgfgraphicnamed{TikZit//compactstructure_capdit}
\InputIfFileExists{TikZit//compactstructure_capdit.tikz}{}{\input{./figures/TikZit//compactstructure_capdit.tikz}}
\endpgfgraphicnamed \qquad\quad %
\beginpgfgraphicnamed{TikZit//compactstructure_cupdit}
\InputIfFileExists{TikZit//compactstructure_cupdit.tikz}{}{\input{./figures/TikZit//compactstructure_cupdit.tikz}}
\endpgfgraphicnamed \qquad\quad %
\beginpgfgraphicnamed{TikZit//compactstructure_snakedit}
\InputIfFileExists{TikZit//compactstructure_snakedit.tikz}{}{\input{./figures/TikZit//compactstructure_snakedit.tikz}}
\endpgfgraphicnamed
 \end{equation}
  \begin{equation}\label{compactstructureslide}
\beginpgfgraphicnamed{TikZit//compactstructure_slidedit}
\InputIfFileExists{TikZit//compactstructure_slidedit.tikz}{}{\input{./figures/TikZit//compactstructure_slidedit.tikz}}
\endpgfgraphicnamed
  \end{equation}
where 
$$%
\beginpgfgraphicnamed{TikZit//anymapdit}
\InputIfFileExists{TikZit//anymapdit.tikz}{}{\input{./figures/TikZit//anymapdit.tikz}}
\endpgfgraphicnamed$$
is an arbitrary diagram in the qufinite $ZX_{\Delta}$-calculus.

The first two diagrams in equation (\ref{compactstructure}) mean the cap $\eta_s$ and the cup $\epsilon_s$ are symmetric, while the last diagram means the connected cap  and cup can be yanked. The first two diagrams of equation (\ref{compactstructureslide}) mean any diagram could move across a line freely, representing the naturality of the swap morphism. The last diagram of equation (\ref{compactstructureslide}) means the swap morphism is self-inverse. Note that now we have a self-dual compact structure rather than a general compact structure, which makes representation of diagrams much easier.

From the rewriting rules noted above, we form a strict self-dual compact closed category $\mathfrak{Z}$ of ZX diagrams. The objects of $\mathfrak{Z}$ are all the positive integers, and the monoidal product on these objects are multiplication of integer numbers. Denote the set of generators listed in Table  \ref{qbzxgeneratordit} as $\mathnormal{G}$. Let $\mathcal{Z}[\mathnormal{G}]$ be a free monoidal category generated by $\mathnormal{G}$ in the following way - i) any two diagrams $D_1$ and $D_2$ are placed side-by-side with $D_1$ on the left of $D_2$ to form the monoidal product on morphisms $D_1 \otimes D_2$, or ii) the outputs of $D_1$ connect with the inputs of $D_2$ when their types all match to each other to form the sequential composition of morphisms $D_2 \circ D_1$. The empty diagram is a unit of parallel composition and the diagram of a straight line is a unit of the sequential composition. Denote the  set of rules listed in Figure \ref{qufiniterules1}, Figure \ref{qufiniterules2}, equations (\ref{compactstructure}) and equations (\ref{compactstructureslide}) by $\mathnormal{R}$.  One can check that rewriting one diagram to another diagram according to the rules of $\mathnormal{R}$ is an equivalence relation on diagrams in $\mathcal{Z}[\mathnormal{G}]$. We also call this equivalence as  $\mathnormal{R}$, then the quotient category $\mathfrak{Z}=\mathcal{Z}[\mathnormal{G}]/\mathnormal{R}$ is  a strict self-dual compact closed category. The qufinite $ZX_{\Delta}$-calculus is seen as a graphical calculus based on the category $\mathfrak{Z}$.

\subsection{Standard interpretation of qufinite $ZX_{\Delta}$-calculus}\label{secc:interpretation}

To ensure that qufinite $ZX_{\Delta}$-calculus is sound, we need to test its rules in a preexisting reliable system which we now describe. These interpretations, however, does not represent the explicit meaning in terms of our consciousness processes. They are given here to test soundness.

Let  $\mathbf{Mat}_{\mathcal{S}}$ be the category whose objects are non-zero natural numbers and whose morphisms $M: m \rightarrow n$ are $n \times m$  matrices taking values in a given commutative semiring $\mathcal{S}$. The composition is matrix multiplication, the monoidal product on objects and morphisms are multiplication of natural numbers and the Kronecker product of matrices respectively. Then $\mathbf{Mat}_{\mathcal{S}}$ is a strict self-dual compact closed category. We give a standard interpretation, namely $\left\llbracket \cdot \right\rrbracket$, for the qufinite $ZX_{\Delta}$-calculus diagrams in $\mathbf{Mat}_{\mathcal{S}}$:
\[
\left\llbracket %
\beginpgfgraphicnamed{TikZit//generalgreenspiderqdit}
\InputIfFileExists{TikZit//generalgreenspiderqdit.tikz}{}{\input{./figures/TikZit//generalgreenspiderqdit.tikz}}
\endpgfgraphicnamed \right\rrbracket=\sum_{i=0}^{d-1}a_j\ket{i}^{\otimes m}\bra{i}^{\otimes n}; a_0=1; a_i\in \mathcal{S};
\]
\[
\left\llbracket %
\beginpgfgraphicnamed{TikZit//redspider0pd}
\InputIfFileExists{TikZit//redspider0pd.tikz}{}{\input{./figures/TikZit//redspider0pd.tikz}}
\endpgfgraphicnamed \right\rrbracket=\sum_{\substack{0\leq i_1, \cdots, i_m,  j_1, \cdots, j_n\leq d-1\\ i_1+\cdots+ i_m\equiv  j_1+\cdots +j_n(mod~ d)}}\ket{i_1, \cdots, i_m}\bra{j_1, \cdots, j_n};
\]
\[
\left\llbracket%
\beginpgfgraphicnamed{TikZit//redclassicd}
\begin{tikzpicture}
	\begin{pgfonlayer}{nodelayer}
		\node [style=none] (0) at (0, 0.75) {};
		\node [style=none] (1) at (0, -0.75) {};
		\node [style=rn] (2) at (0, 0) {};
		\node [style=none] (3) at (0.25, 0) {$d_j$};
	\end{pgfonlayer}
	\begin{pgfonlayer}{edgelayer}
		\draw (0.center) to (1.center);
	\end{pgfonlayer}
\end{tikzpicture}}
\endpgfgraphicnamed\right\rrbracket= \sum_{i=0}^{d-1}\ket{i}\bra{i\oplus j};\quad
  \left\llbracket%
\beginpgfgraphicnamed{TikZit//triangled}
\begin{tikzpicture}
	\begin{pgfonlayer}{nodelayer}
		\node [style=none] (0) at (0, 0.5) {};
		\node [style=none] (1) at (0, -0.5) {};
		\node [style=triangle] (2) at (0, 0) {};
		\node [style=none] (3) at (-0.25, 0) {$d$};
	\end{pgfonlayer}
	\begin{pgfonlayer}{edgelayer}
		\draw (0.center) to (1.center);
	\end{pgfonlayer}
\end{tikzpicture}}
\endpgfgraphicnamed\right\rrbracket=\ket{0}\bra{0}+\sum_{i=1}^{d-1}(\ket{0}+\ket{i})\bra{i}; \quad
\left\llbracket%
\beginpgfgraphicnamed{TikZit//idqudit}
\begin{tikzpicture}
	\begin{pgfonlayer}{nodelayer}
		\node [style=none] (0) at (0, 0.5) {};
		\node [style=none] (1) at (0, -0.5) {};
		\node [style=none] (2) at (0.25, -0.5) {$d$};
	\end{pgfonlayer}
	\begin{pgfonlayer}{edgelayer}
		\draw (0.center) to (1.center);
	\end{pgfonlayer}
\end{tikzpicture}}
\endpgfgraphicnamed\right\rrbracket= \sum_{i=0}^{d-1}\ket{i}\bra{i}; 
   \]
\[
 \left\llbracket%
\beginpgfgraphicnamed{TikZit//binderdit}
\InputIfFileExists{TikZit//binderdit.tikz}{}{\input{./figures/TikZit//binderdit.tikz}}
\endpgfgraphicnamed\right\rrbracket= \sum_{k=0}^{s-1}\sum_{l=0}^{t-1}\ket{kt+l}\bra{kl}; 
 \quad \quad
  \left\llbracket%
\beginpgfgraphicnamed{TikZit//binderditflip}
\InputIfFileExists{TikZit//binderditflip.tikz}{}{\input{./figures/TikZit//binderditflip.tikz}}
\endpgfgraphicnamed\right\rrbracket= \sum_{k=0}^{st-1}\ket{[\frac{k}{t}]}\ket{k-t[\frac{k}{t}]}\bra{k}; \quad\quad   \left\llbracket%
\beginpgfgraphicnamed{TikZit//emptysquare}
\InputIfFileExists{TikZit//emptysquare.tikz}{}{\input{./figures/TikZit//emptysquare.tikz}}
\endpgfgraphicnamed\right\rrbracket=1;
 \]

\[
 \left\llbracket%
\beginpgfgraphicnamed{TikZit//swapd}
\InputIfFileExists{TikZit//swapd.tikz}{}{\input{./figures/TikZit//swapd.tikz}}
\endpgfgraphicnamed\right\rrbracket= \sum_{k=0}^{s-1}\sum_{l=0}^{t-1}\ket{kl}\bra{lk}; 
 \quad
  \left\llbracket%
\beginpgfgraphicnamed{TikZit//capdit}
\begin{tikzpicture}
	\begin{pgfonlayer}{nodelayer}
		\node [style=none] (0) at (0.5, -0.25) {};
		\node [style=none] (1) at (-0.5, -0.25) {};
		\node [style=none] (2) at (-0.75, -0.25) {$s$};
		\node [style=none] (3) at (0.75, -0.25) {$s$};
	\end{pgfonlayer}
	\begin{pgfonlayer}{edgelayer}
		\draw [in=90, out=90, looseness=1.75] (1.center) to (0.center);
	\end{pgfonlayer}
\end{tikzpicture}}
\endpgfgraphicnamed\right\rrbracket= \sum_{i=0}^{s-1}\ket{i}\ket{i};  \quad
   \left\llbracket%
\beginpgfgraphicnamed{TikZit//cupdit}
\begin{tikzpicture}
	\begin{pgfonlayer}{nodelayer}
		\node [style=none] (0) at (0.75, 0.25) {$s$};
		\node [style=none] (1) at (-0.75, 0.25) {$s$};
		\node [style=none] (2) at (-0.5, 0.25) {};
		\node [style=none] (3) at (0.5, 0.25) {};
	\end{pgfonlayer}
	\begin{pgfonlayer}{edgelayer}
		\draw [in=-90, out=-90, looseness=1.75] (2.center) to (3.center);
	\end{pgfonlayer}
\end{tikzpicture}}
\endpgfgraphicnamed\right\rrbracket= \sum_{i=0}^{s-1}\bra{i}\bra{i};
      \]

\[  \llbracket D_1\otimes D_2  \rrbracket =  \llbracket D_1  \rrbracket \otimes  \llbracket  D_2  \rrbracket; \quad 
 \llbracket D_1\circ D_2  \rrbracket =  \llbracket D_1  \rrbracket \circ  \llbracket  D_2  \rrbracket;
  \]
where 
$s, t \in \mathbb N \backslash\{0\};   \bra{i} =\overbrace{(\underbrace{0,\cdots,1}_{i+1}, \cdots, 0)}^{d}; ~ \ket{i}=(\overbrace{(\underbrace{0,\cdots,1}_{i+1}, \cdots, 0)}^{d})^T;  i \in \{0, 1,\cdots, d-1\}$;  and $[r]$ is the integer part of a real number $r$.

One can verify that the qufinite $ZX_{\Delta}$-calculus is sound in the sense that for any two diagrams $D_1, D_2 \in  \mathfrak{Z}$, $ D_1= D_2$ must imply that $\llbracket D_1  \rrbracket =  \llbracket  D_2  \rrbracket$. This standard interpretation $\llbracket \cdot \rrbracket$ is actually a strict symmetric monoidal functor from $\mathfrak{Z}$ to $\mathbf{Mat}_{\mathcal{S}}$.

According the standard interpretation, 
if $S$ is the field of complex numbers, then the green spider corresponds to the computational basis  $\ket{i}\}_{i=0}^{d-1}$, with $d-1$ phase angles. The red spider corresponds to the Fourier basis coming from Fourier transformation of the computational basis, up to a global scalar. The red $d_j$ diagram represents the $j$-th unitary which is also a permutation matrix, with $j$ ranging from 0 to $d$.  The triangle diagram labelled with $d$ acts as a successor of phase parameters (adding 1’s to them). The two trapezium diagrams represent unitaries between the Hilbert space of $H_s$ $\otimes H_t$  and the Hilbert space $H_{st}$, these two diagrams are invertible to each other.

\subsection{The perceived division of Alaya Consciousness}
\label{secc:interpretation2}

Now, we model the perceived division of Alaya consciousness. As we have introduced in section \ref{secc:YPC}, the content of the perceived version of Alaya consciousness is the phenomenon of the physical world and the body which is supposed to have the same mathematical structure for all sentient beings in this world (not necessarily the same types, which may bring specificity and a treatment for individuality). Since each physical object is supposed to be composed of quantum systems, the perceived version of Alaya consciousness is modelled here by the category  $\mathbf{FdHilb}$: the category whose objects are all finite dimensional complex Hilbert spaces  and whose morphisms are linear maps between the Hilbert spaces with ordinary composition of linear maps as compositions of morphisms. The usual Kronecker tensor product is the monoidal tensor, and the field of complex numbers $\mathbb{C}$ (which is a one-dimensional Hilbert space over itself) is the tensor unit.  $\mathbf{FdHilb}$ is the category of quantum processes which composes the physical world. Since the body is a part of the physical world, the body part of the perceived division of alaya consciousness may be modelled by a subcategory of $\mathbf{FdHilb}$.

\subsection{The perceiving division of Alaya consciousness}
 
The function of the perceiving division of Alaya consciousness is to perceive the perceived division, which means a perceiving action from the subject (perceiving) to the object (perceived) of the Alaya consciousness. Since a functor is a structure preserving map or transformation from one category to another one, our first attempt is to model the perceiving division of Alaya consciousness by a functor from $ \mathfrak{Z}$ to $\mathbf{FdHilb}$. This functor is set up as a modification of the standard interpretation functor $\llbracket \cdot \rrbracket$, i.e.: just choose a semiring homomorphism $f$ from $\mathcal{S}$ to $\mathbb{C}$ and let $\{ \ket{i}\}_{i=0}^{d-1}$ a standard basis of a Hilbert space with dimension $d$, then replace $a_i$ with $f(a_i)$ in the codomain of the interpretation $\llbracket \cdot \rrbracket$. One can check that a monoidal functor is obtained in this way, where a semiring homomorphism from $\mathcal{S}$ to $\mathbb{C}$ is selected.

\section{The Unity of Experience}
\label{secc:Unity}

As a consequence of our first simple approach to model conscious experience, we consider the combination problem of the unity of experience. 

We suggest that some aspects of Alaya consciousness can be modelled by qufinite  $ZX_{\Delta}$-calculus, making it a serious and somewhat justified attempt. A general diagram represents some  primary conscious experience and a diagram with outputs but without inputs represent a state of consciousness. Sequential composition of two diagrams represents two successive conscious processes happening one after the another, while parallel composition of two diagrams represents two conscious processes happening simultaneously. These processes may compound to generate more complex experiences.

Our approach is an alternative to conserve the irreducible and fundamental nature of experience. It is not, however, the only one. Panpsychism and Panprotopsychism, among others philosophies, also consider experience seriously, but these two assigns a quantifiable character to that experience. According to these views, consciousness is present in all fundamental physical entities \cite{Chalmers2013a} and the composition of basic blocks of experience creates our conscious experience. 

Nevertheless, an important question remains for those irreductible attemps: How "microphenomenal seeds of consciousness" constitute macrophenomenal conscious experiences as we experience them? ---the so-called combination problem for Panpsychism and Panprotopsychism \cite{Chalmers2016}. This problem convey an specific form of the hard problem of consciousness, i.e. how could the key features of consciousness, like its unity, arise from any kind of interactions of physical atomic experiences (given by physical theory of your choice), no matter how many of these atoms and how complicated the interactions are, which have none of those key features? \cite{ThomasNagel1974, Searle2000, Chalmers1995}. In other words, how these building blocks of experience compound one single unified macro phenomenal subjective experience \cite{Bayne2012}: the phenomenal unity of experience \cite{Revonsuo1999, Bayne2012}. In Panpsychism and Panprotopsychism, the dualism between mind and matter is now replaced by two modes, micro and macro experience, of the same ontology.

Remarkable, the combination problem has three aspects \cite{Chalmers2016}: structural, subject and quality. Each one of these aspects leads to a specific sub-problem. On the one hand, the structure of the micro world, mostly associated with quantum mechanics, gives the impression of being different from the structure of macro experiences. This is the structural mismatch problem, which also appears between macro experience structure and macro physical structures in the brain \cite{Chalmers2016}. On the other hand, there is the question of how micro subject combine to give rise to macro subjects, and how micro qualities combine to give macro qualities. It seems that no group of micro subjects need the existence of a macro subject, and additionally, it is not clear how possible limited micro qualities yield to the many macro qualities that can be experienced, including different colors, shapes, sounds, smells, and tastes (for detail see \cite{Chalmers2016}). According to Chalmers, a satisfactory solution of the combination problem must face all these three aspects.

Our framework targets all of these aspects of the combination problem. First, the mathematical structure of the qufinite $ZX_{\Delta}$-calculus for Alaya consciousness is a unification of all dimensional qudit ZX-calculus. If generators are interpreted in Hilbert space, the latest becomes a graphical language for quantum theory. This means that the $ZX_{\Delta}$-calculus for conscious processes shares a similar structure to quantum theory \footnote{Please note that a similar structure means similar mathematical relationships. Two very different phenomena in \textit{nature} may share the same structure and being modelled by the same equations, or more generally, the same categories.}. This similarity solves the mismatch at the level of micro experience, without any ontological commitment to quantum particles (e.g. different to Hameroff and Penrose model \cite{Hameroff2014}). At the level of macro experiences we avoid any match or mismatch with macro physical structures because the model does not reduce experience to neural events (non-isomorphic relationship). It means that conscious experience does not need to share the same structure that \textit{classical} neurons. Second, the model does not distinguish between subject and quality, everything is a conscious process, a conscious experience. Those fundamental conscious processes of reality, namely the generators of the theory, compound other conscious processes just by means of connecting them together: via sequential and parallel composition. The result of those compositions are other more complex subjective and qualitative processes. New compounded processes depend on the basic generators, while the generators are interrelated to define themselves via rewriting rules (axioms), representing more complex experiential relationships; i.e. each process need other processes to specify itself. 

In our framework, unity of consciousness is naturally described as a result of process composition \cite{Signorelli2020f}. If someone insists on generators being matched with subjects or agents, then micro (generators) and macro subjects (composition of generators) necessitate themselves as imposed by the co-dependent nature. This deals with the problem of subject/quality composition at the level of Alaya consciousness (please check \cite{Chalmers2016} for details). This treatment also allows to deal with the combination of specific qualities as the result of compositions in the seventh mental consciousness, work that form part of an ongoing project.  

Summarizing, while in Panpsychism the division between micro and macro is given by physical systems (e.g. atoms versus neurons, or neurons versus neural assembles), in our framework there is no such distinction. The distinction just vanish, since we consider generators that already carry the properties of the whole (experience), following the compositionality principle. The choice of generators might seems arbitrary, and it is, as far as we do not have a sound and complete rewriting system for conscious experience.

\section{Conclusions}
 \label{secc:Con}
 
Our framework is based on arbitrary commutative semirings as a compositional model of consciousness, with the emphasis on its potential use for the mathematical and structural studies of consciousness \cite{Yoshimi2007, Prentner2019, Tsuchiya2020}. We introduced generators and processes as abstract mathematical structures to target some aspects of the conscious experience which are independent of their physical realizations. This introduction is inspired by the Yogacara school of Buddhism and other philosophies assuming that consciousness is primary or fundamental. In this first attempt, we have focused on Alaya consciousness, its co-dependent character and its perceived and perceiving division. It allow us to make a first approach to the dual question of how the objective world \textit{emerges} from the subjective one. 

A future approach may target more details on the internal structure of Alaya consciousness, taking for example, six features in formal analogy with the seed metaphor from \cite{TatWei}. Moreover, we leave for future work the model of mental, manas and the five sense-consciousnesses. In the future, we also expect to generalise the qufinite $ZX_{\Delta}$-calculus to the infinite dimensional case, from which standard quantum mechanics might be recovered. It is to be noted that we have not recovered standard quantum mechanics. To do so would mean generalising our model to derive the standard quantum mechanics described by the Schr\"{o}dinger equation. This is important in order to give a definitive answer to the dual problem of consciousness introduced above. 

Other very interesting models also aim to target that question. For example, the conscious agent model intends to recover fundamental physics from the agent’s interactions, as for instance in quantum mechanics \cite{Hoffman2014}. Sadly, it is not clear that current versions of the conscious agent model are capable of recovering the entire objective realm (see objections and replies section in \cite{Hoffman2014}). In our framework part of the reconstruction goal pursued by the conscious agent model is achieved for free, without overhead, invoking only the simple structure of SMC in relationship with phenomenal aspects. In doing so, our approach to consciousness processes and quantum theory share a similar, but not the same, mathematical structure.

It allows a compositional treatment of the combination problem of basic experience that may give rise to complex ones. One very influential model of consciousness also attempting to target that question is the the integrated information theory (IIT) \cite{Oizumi2014}. This model, however, conveys a Panpsychist view. Unfortunately, the model is not neutral about the physical neural substrate, and although it intends to highlight the primacy of consciousness, in practice, the current version falls in reductive accounts. Such models also claim compositions, but they are not compositional in the sense exposed here. In IIT, the minimal elements of the theory are gates that are not conscious, while consciousness emerges from the right causal combination of these gates (integration). Contrary, compositionality in our sense means that the minimal compositional elements, i.e. generatores, represents already conscious experiences.

Our model, the conscious agent model and IIT, all share the same goal of \textit{mathematize phenomenology}. Although with different philosophical commitments, the common point of departure is that axioms and postulates consider aspects of consciousness as primary. However, we close by remarking that a process theory for consciousness is not only about modelling consciousness with any type of mathematics (as other models), but about modelling consciousness with category theory in a graphical form, i.e. axiomatic mathematics. This form of mathematics explicitly introduces structures, assumptions and axioms, plus the possibility of compositional treatments. We believe that because being foundational, this approach is better suited to describing the conscious experience as fundamental. Finally, we are hopeful that due to its co-dependent feature, and sufficient generality, our framework may pave the way for further research on the scientific study of conscious experience and its phenomenology.

\vspace{6pt} 


\authorcontributions{ Conceptualization, CMS and QW; investigation CMS, QW and IK; writing-original draft preparation, CMS; writing-review and editing, CMS, QW and IK; visualization, CMS and QW.}

\funding{CMS is funded by Comisión Nacional de Investigación Ciencia y Tecnología (CONICYT, current ANID) through Programa Formacion de Capital Avanzado (PFCHA), Doctoral scholarship Becas Chile: CONICYT PFCHA/DOCTORADO BECAS CHILE/2016 - 72170507. QW was supported by AFOSR grant FA2386-18-1-4028.}

\acknowledgments{The authors appreciate valuable feedback and discussions from Bob Coecke, Konstantinos Meichanetzidis and Robert Prentner. The authors would also like to thank the anonymous reviewers and editors for their helpful comments.}

\conflictsofinterest{The authors declare no conflict of interest.} 




\reftitle{References}






\externalbibliography{yes}
\bibliography{library}



\end{document}